\documentclass
[showkeys,superscriptaddress,10pt,balancelastpage,nofootinbib,
nobibnotes,fleqn]{revtex4}%
\usepackage{amsfonts}
\usepackage{amsmath}
\usepackage{amssymb}
\usepackage{graphicx}%
\providecommand{\U}[1]{\protect\rule{.1in}{.1in}}
\ifx\pdfoutput\relax\let\pdfoutput=\undefined\fi
\newcount\msipdfoutput
\ifx\pdfoutput\undefined\else
\ifcase\pdfoutput\else
\ifx\paperwidth\undefined\else
\ifdim\paperheight=0pt\relax\else\pdfpageheight\paperheight\fi
\ifdim\paperwidth=0pt\relax\else\pdfpagewidth\paperwidth\fi
\fi\fi\fi
\begin{document}
\title{Atom Optics Quantum Pendulum}
\author{Muhammad Ayub \thanks{ayubok@yahoo.com}}
\affiliation{Department of Electronics, Quaid-i-Azam University, 45320, \
Islamabad, Pakistan.}
\affiliation{Theoretical Plasma Physics Division, PINSTECH, Nilore, Islamabad,
Pakistan.}
\author{Khalid Naseer}
\affiliation{Department of Electronics, Quaid-i-Azam University, 45320, \
Islamabad, Pakistan.}
\affiliation{Department of Physics, University of Sargodha, Sargodha, Pakistan.}
\author{Manzoor Ali}
\affiliation{Department of Electronics, Quaid-i-Azam University, 45320, \
Islamabad, Pakistan.}
\affiliation{Department of Physics, Karakurum International University Gilgit,
Pakistan.}
\author{Farhan Saif\thanks{fsaif@yahoo.com}}
\affiliation{Department of Electronics, Quaid-i-Azam University, 45320, \
Islamabad, Pakistan.}
\date{\today}

\begin{abstract}
We explain the dynamics of cold atoms, initially trapped and cooled in a
magneto-optic trap, in a monochromatic stationary standing electromagnetic
wave field. In the large detuning limit the system is modeled as a nonlinear
quantum pendulum. We show that wave packet evolution of the quantum particle
probes parametric regimes in the quantum pendulum which support classical
period, quantum mechanical revival and super revival phenomena.
Interestingly, complete reconstruction in particular parametric regime at
quantum revival times is independent of potential height.

\end{abstract}

\maketitle

\section{Introduction}

Pendulum in quantum mechanics \cite{Condon} is a subject of great interest
when the question comes to explain hindered internal rotations in chemistry 
\cite{Lister}, quantum features of scattering atoms in quantum optics
\cite{Dyrting, Chu}, perturbation theory methods to study weak field effects in
quantum mechanics \cite{Schwartz,Jhonston, Kiang}, dynamics of Bose-Einstien
condensates in optical lattices for small nonlinearity\cite{H.Pu} and many
other physical systems. Comprehensive study of quantum pendulum has brought
to light its various aspects which include the structure of energy spectrum,
the time evolution focusing on the quantum revivals
\cite{WRevivals,Robinett,Doncheski,Saif} and asymptotic Mathieu solutions using
algebraic methods\ \cite{Frenkel}.

Super cold atoms in optical standing field is an area of both theoretical
and experimental interest and modeled as quantum pendulum \cite{Dyrting}.
The system is of great importance in classical and quantum domain as an
explicit time dependence in phase or amplitude makes the classical
counterpart chaotic.

Bose-Einstein condensates in an optical lattice \cite{Morsch} is a mile
stone and an important paradigm to study atomic condensate evolution, and
its phase transition from Mott insulator to superfluid state \cite{MAP
Fisher,Wright,Jaksch,Greiner,BlochJPB,Eckardt,Bloch}. Quantum revivals in
the system present a profound manifestation of quantum interference. The
phenomenon occurs as optical lattice potential is\ not perfectly harmonic
and level spacing varies with quantum number. The revivals provide useful
information about the coherence time of the atoms in the optical lattice.

In \cite{Robinett,Doncheski} energy spectrum and time scales encoded in the
spectrum are discussed in detail. In this contribution, we extend the work
following the same approach and explain the evolution of super cold weakly
condensed atoms in time independent\ optical lattice. Furthermore, we
explain the energy spectrum and eigen states of the system both analytically
and numerically. The time dynamics of the wave packet also translates the
relevance of energy spectrum with classical phase space. We probe
distribution of energy eigen states of the potential via wavepacket
evolution. For the simplicity we limit our discussion to deep optical
lattice where tunneling does not play significant role.\ We show: (i) An
equally spaced local energy spectrum deep in the well which reflects itself
in the reconstruction of the wavepacket at classical periods; (ii) Beyond
this region, the behavior changes as the level spacing modifies itself. The
nonlinear behavior dominates and controls the dynamics as we go farther from
the deep-in-the-well condition; (iii) Numerical calculations with enhanced
efficiency and accuracy in the presence of analytical relationships explain
the detailed dynamics of the quantum pendulum and its parametric dependence
on anharmonicity.

In section II, we discuss atomic interaction with optical lattice and obtain
quantum pendulum. In section III, we work out analytical solutions of the
potential. In section IV, simplified quantum pendulum is modeled as series
expansion of cosine potential and effect of each term of the series is
explained which plays the effective role in the atomic evolution$\ $in
quantum pendulum. Results are discussed in section V.

\section{The Model}

We consider super cold two-level atoms interacting with a classical
monochromatic standing wave field, $E(x,t)=\hat{e}_{y}[\varepsilon _{o}\cos
(k_{L}x)e^{-i\omega t}+c.c.].$ Here, $\omega $ and $k_{L}$ are,
respectively, the frequency and the wave number of the laser field, and $%
\hat{e}_{y}$ is the polarization vector. We assume that the cavity end
mirrors, in $yz$-plane, reflect the incoming light wave along $x$-axis and,
therefore, determine the position of the nodes of the standing light along
the axis. In dipole and rotating-wave approximations the atom-field
interaction is controlled by the Hamiltonian, 
\begin{equation}
H=\frac{p_{x}^{2}}{2M}+\hbar \omega _{o}|e\rangle \langle e|-\{\overset{%
\rightharpoonup }{d}.\hat{e}_{y}\varepsilon _{o}\cos (k_{L}x)\times
e^{i\omega t}\sigma _{+}+H.c\}.  \label{hal}
\end{equation}%
Here, $\hbar \omega _{o}$ is the energy difference between atomic excited
state, $|e\rangle ,$\ and ground state, $|g\rangle ,$ and $\overset{%
\rightharpoonup }{d}$\ indicates atomic dipole moment. Moreover, $p_{x}$ is
the center-of-mass momentum of the atom along the cavity axis, $M$ is the
atomic mass and $\sigma _{\pm }$ are the raising and lowering operators
defined as; $\sigma _{+}=|e\rangle \langle g|$ and $\sigma _{-}=|g\rangle
\langle e|$. Furthermore, we have considered condensate densities so low
that interparticle interactions are negligible and single particle approach
remain applicable \cite{Gardiner,Eckardt2009}. Thus, we represent the wave
function of the system at any time of interaction, $t,$ as, $|\psi \left(
x,t\right) \rangle =\psi _{g}\left( x,t\right) |g\rangle +\psi _{e}\left(
x,t\right) e^{-i\omega t}|e\rangle $. In the presence of sufficiently large
detuning between the atomic transition frequency and the field frequency,
i.e., $\delta _{L}=\omega _{0}-\omega ,$ allows us to neglect spontaneous
emission. Furthermore, the same consideration allows us to eliminate the
excited state adiabatically and leads to effectively describe the evolution
of the atom within the electromagnetic field in its ground state. Hence the
dynamics of an atom, is governed effectively by the Hamiltonian,

\begin{equation*}
H=\frac{p_{x}^{2}}{2M}-\frac{\hbar \Omega _{eff}}{8}\cos (2k_{L}x).
\end{equation*}%
Here, $\Omega _{eff}=\Omega ^{2}/\delta _{L}$ is the effective Rabi
frequency, where, $\Omega =\overset{\rightharpoonup }{d}.\hat{e}%
_{y}\varepsilon _{o}/\hbar $ is Rabi frequency. The probability to find an
atom, therefore, in the excited state is negligible as a consequence of
large detuning, the evolution properties of the atom in the field are
completely determined by the ground state amplitude. It is useful to define
the following dimensionless quantities $t=\omega t,$ $\bar{x}=2k_{L}x,$ $%
p=(2k_{L}/M\omega )p_{x}$ and $\bar{H}=H(4k_{L}^{2}/M\omega ^{2}).$\ We get
the dimensionless effective Hamiltonian as%
\begin{equation}
\bar{H}=\frac{p^{2}}{2}-V_{0}\cos \bar{x},  \label{a}
\end{equation}%
where, the parameters $V_{0}=\epsilon \Omega _{eff}/\omega ^{2}$ is the
effective potential depth and $\epsilon =\hbar k_{L}^{2}/2M$ is the recoil
shift. In case $\delta _{L}\,>0$ and the laser is tuned red to the atomic
transition, we find $V_{0}$ as positive. Hence, a phase difference between
the potential and laser intensity shifts the location of the potential
minima such that they coincide with the locations of the laser intensity
maxima. The atom is, therefore, attracted towards the intensity maxima. In
the other case, that is $\delta _{L}<0,$ the phase difference between the
potential and laser intensity disappears and the atom is attracted towards
intensity minima. The quantized system has another controlling parameter as
scaled Planck's constant $k^{\hspace{-2.1mm}-}=8\epsilon /\omega ,$ which
follows the commutation relation $i[p,\bar{x}]=k^{\hspace{-2.1mm}-}.$ Here, $%
\bar{H}$ effectively defines the dynamics of a quantum particle as a quantum
pendulum.

The eigen states for quantum rotor have the spatial periodicity condition $%
\phi _{n}(\bar{x})=\phi _{n}(\bar{x}+2\pi ),$ while, eigen states for an
atom in optical lattice obeys the general Bloch condition $\phi _{n}(\bar{x}%
)=e^{i2\pi \nu }\phi _{n}(\bar{x})$ where $\nu $ indicates the
quasi-momentum of the eigen state. The general Bloch condition satisfy the
spatial boundary condition for quantum rotor, when $\nu $ has integer
values. In contrast, the range of $\nu $ is continuous for optical lattices.
In both cases the potential only couples the eigen states on a momentum
ladder where the ladder spacing is $2\hbar k_{L}.$ However, for optical
lattices, there are many momentum ladders independent from each other while
for quantum rotor there is only one, centered at $p=0$ \cite{Madison}.

\section{Schr\"{o}dinger Equation for Quantum Pendulum and Mathieu Equation}

The dynamics of the atom interacting with a standing wave field in the large
detuning limit is effectively controlled by the time independent Schr\"{o}%
dinger wave equation, expressed as%
\begin{equation*}
-\frac{k^{\hspace{-2.1mm}-2}}{2}\frac{\partial ^{2}\psi (\bar{x})}{\partial
x^{2}}-V_{0}\cos (x)\psi (\bar{x})=E_{n}\psi (\bar{x}),
\end{equation*}%
where, for simplicity we write $\psi _{g}=\psi $ and $\bar{x}\rightarrow x$.
We rewrite the position variable $x$\ as $x\equiv x-\pi $, and get Mathieu
equation, 
\begin{equation*}
\frac{\partial ^{2}\psi _{n}(x)}{\partial x^{2}}+[a_{n}-2q\cos (x)]\psi
_{n}(x)=0,
\end{equation*}%
which states that for any given $a_{n}$ and $q,$ we have a series of
solution, $\psi _{n}(x),$ for the differential equation, labeled by index $%
n. $ Hence, $\psi _{n}(x)$ defines the eigen function of the system and 
\begin{equation}
q\equiv \frac{V_{0}}{k^{\hspace{-2.1mm}-2}}=\frac{\hbar \Omega _{eff}}{%
64E_{R}}\ \text{and\ }a_{n}\equiv \frac{2E_{n}}{k^{\hspace{-2.1mm}-2}},
\label{q}
\end{equation}%
express, respectively, effective potential depth and Mathieu characteristic
parameter leading to eigen energies.

Atom in an optical lattice may observe deep or shallow potential depths
corresponding to its energy. Following equation \ref{q}, we scale effective
potential depth by recoil energy of an atom, $E_{R}=\hbar \epsilon $. The
atom in cosine potential observes a deep optical potential if effective
potential depth $V_{0}$ is of the order of a few hundred single photon
recoil energies and temperature of the atom is around recoil temperature. In
this case the dynamics of the quantum particle in the individual well is
independent and one obtains multiple realization of anharmonic oscillators.
On the other hand when the depth of the potential is just few recoil
energies and atom is at about recoil temperatures, \ it sees a shallow
potential. In this case, the quantum mechanical effects caused by spatial
periodicity of optical lattices such as formation of Bloch waves become
important. Furthermore the levels are broadened into bands due to resonant
tunneling between adjacent wells \cite{Holthaus}. Tunneling in the low-lying
bands is suppressed as the well depth is increased and particle motion is
dominated by single-well dynamics as it is discussed for large $V_{0}$.

A moderate values of $V_{0}$, with an effective Planck's constant of order
unity, indicates the deep quantum regime. The semiclassical dynamics of the
atom in the standing wave field are observed however for large values of $%
V_{0}$, which correspond to small values of effective Planck's constant $k^{%
\hspace{-2.1mm}-}$ and here we find several tightly bound energy bands.
Quantum mechanical effects for small $V_{0}$, become important once the
atomic de Broglie wavelength $\frac{2\pi \hbar }{P}$ significantly exceeds
the lattice constant $d=\frac{1}{2}\frac{2\pi }{k_{L}}=\frac{\lambda _{L}}{2}%
.$ This gives the condition%
\begin{equation*}
\frac{P^{2}}{2M}\ll \frac{4\hbar ^{2}k_{L}^{2}}{2M}\equiv 4E_{R}.
\end{equation*}

For a fixed value of $q$, there are countably infinite number of solutions,
labeled by $n.$ However, only for specific characteristic values of the
parameter $a_{n}$, the solutions will be periodic, with periods $\pi $ or $%
2\pi $ in the variable $x,$ which are denoted by $a_{n},$ and $b_{n},$
respectively for the even and odd solutions. Because of the intrinsic parity
of the potential, the solutions can be characterized as being even, $%
ce_{n}(x,q)$ or cosine-like for integral values of $n$, with $n\geq 0.$
Whereas they are odd, $se_{n}(x,q)$ or sine-like for integral values of $n$,
with $n\geq 1$. Limiting cases i.e.$.$ $q=0,$ $q\ll 1$ and $q\gg 1$ for
quantum pendulum are discussed in detail in Reference \cite{Doncheski}.

Approximate expressions for the characteristic values of $a_{n}$ and $b_{n}$
both in $q\ll 1$ and in $q\gg 1$ limits are provided by references \cite%
{AbramowitzStegun, McLachlan}. For the limit, $q\ll 1$, we find that the $%
a_{n}$, $b_{n}$ are approximately degenerate for $n\gtrsim 7,$ that is,%
\begin{equation}
a_{n}\simeq b_{n}=n^{2}+\tfrac{q^{2}}{2(n^{2}-1)}+\tfrac{(5n^{2}+7)q^{4}}{%
32(n^{2}-1)^{3}(n^{2}-4)}+...\text{.}  \label{eq2}
\end{equation}%
The above expression is not limited to integral value of $n$ and is a very
good approximation when $n$ is of the form, $m+\frac{1}{2}$. In case of
integral value of $n=m,$ the series holds only up to the terms not involving 
$n^{2}-m^{2}$\ in the denominator. The difference between the characteristic
values for even and odd solutions satisfy%
\begin{equation*}
a_{n}-b_{n}=O(\tfrac{b_{n}}{n^{n-1}})\text{ \ \ \ \ as }n\rightarrow \infty
\end{equation*}

In other limiting case, when $q\gg 1$ and the spectrum is oscillator like,
we find%
\begin{equation}
a_{n}\approx b_{n+1}\approx -2q+2s\sqrt{q}-\tfrac{s^{2}+1}{2^{3}}-\tfrac{%
s^{3}+3s}{2^{7}\sqrt{q}}-..........,  \label{eq3}
\end{equation}%
where, $s=2n+1.$ It has, thus, $(n+\frac{1}{2})k^{\hspace{-2.1mm}-}\omega
_{h}$ dependence in lower order, which resembles harmonic oscillator energy
for $\omega _{h}=2\sqrt{V_{0}}$. Here in the deep optical lattice limit, the
band width is defined as \cite{AbramowitzStegun}%
\begin{equation}
b_{n+1}-a_{n}\simeq \tfrac{2^{4n+5}\sqrt{\frac{2}{\pi }}q^{\frac{n}{2}+\frac{%
3}{4}}\exp (-4\sqrt{q})}{n!},  \label{bandwith}
\end{equation}%
which shows that in the deep optical lattice ($q\gg 1$\ limit) energy bands
are realized as degenerate energy levels as the band width is negligible.
The band structure of optical lattice is shown in figure-1, for large $q,$\
near the bottom of the lattice,thin band are seen, band width increases and
band gap decreases as we move towards the top of the lattice potential well.

As a consequence, we suppress atomic tunneling in deep optical lattice
limit.\ The hoping matrix element, $J,$\ explain the tunneling between
adjacent sites for deep optical lattice \cite{Jaksch,BlochJPB,Bloch}, viz,%
\begin{equation}
J=\frac{4}{\sqrt{\pi }}E_{R}(V_{o}/E_{R})^{\frac{3}{4}}\exp (-2\sqrt{%
V_{o}/E_{R}}).  \label{Tunneling Matrix}
\end{equation}%
Equations \ref{bandwith}, \ref{Tunneling Matrix} show that the width of the
bands corresponds to tunneling of the atom from one lattice site to the
other. In the limit of deep lattice potentials, this probability will be
exponentially small and band width therefore, reduces exponentially as a
function of lattice potential depth.

Using algebraic methods, asymptotic even mathieu functions are \cite{Frenkel}
given as

\begin{gather}
ce_{n}(x,q)=\tfrac{%
(n^{4}-6n^{3}+11n^{2}-6n)D_{n-4}+(4n-4n^{2})D_{n-2}-4D_{n+2}-D_{n+4}}{64%
\sqrt{q}}  \notag \\
+\frac{1}{1024q}[\frac{1}{8}%
(n^{8}-28n^{7}+322n^{6}-1960n^{5}+6769n^{4}-13132n^{3}  
+13068n^{2}-5040n)D_{n-8}-(n^{6}-15n^{5}+85n^{4}  \notag \\
-225n^{3}+274n^{2}-120n)D_{n-6}+4(n^{5}-7n^{4}+17n^{3}  
-17n^{2}+6n)D_{n-4}-(n^{4}+26n^{3}-37n^{2}+10n)D_{n-2}+  \notag \\
(-36-25n+n^{2})D_{n+2}-4(n+2)D_{n+4}+D_{n+6}+\frac{1}{8}D_{n+8}]  
+D_{n}+O(\frac{1}{q^{\frac{3}{2}}}),
\end{gather}%
where, in terms of Hermit polynomials $D_{n}(\alpha )=\frac{1}{2^{\frac{n}{2}%
}}\exp (-\frac{\alpha ^{2}}{4})H_{n}(\frac{\alpha }{\sqrt{2}})$and $\alpha $
is defined as $\alpha =2q^{\frac{1}{4}}\cos (x).$ The normalization factor
for $ce_{n}(x,q)$ up to order $O(\frac{1}{q^{2}})$ is%
\begin{gather}
\tfrac{1}{C_{n}^{2}}=\tfrac{\sqrt{2}n!}{\sqrt{\pi } q^{\tfrac{1}{4}}}[1+%
\tfrac{2n+1}{8(\sqrt{q})}+\tfrac{n^{4}+2n^{3}+263n^{2}+262n+108}{2048q} 
+\tfrac{6n^{5}+15n^{4}+1280n^{3}+1905n^{2}+1778n+572}{16384q^{\frac{3}{2}}}%
+o(\tfrac{1}{q^{2}})],
\end{gather}%
here eigen states are normalized to $\pi .$

\section{Bound States of an Optical Lattice}

Around the minima of lattice sites harmonic evolution prevails and in the
presence of the higher order terms it gradually modified to the original
potential. Microscopic investigation of the atom-optical system, using term
by term contribution of cosine potential expansion reveals the dominant role
of the system's particular parametric regime in the formation of
eigen-states and eigen energies. This leads to simplified analytical
solutions around the potential minima in the system, as discussed below.

\subsection{\textbf{Harmonic Oscillator Like Limit.}}

In the deep lattice limit, the potential near the minima can be approximated
as quadratic. Thus the particle placed near the minima of the cosine
potential, experiences a harmonic potential. The energy in this regime is
obtained as, $E_{n}^{(0)}=(2n+1)k^{\hspace{-2.1mm}-}\sqrt{V_{0}}-V_{0},$
which can be identified in equation \ref{eq3}, by ignoring square and higher
powers in $s.$ The eigen states of quadratic potential are given as $\phi
_{n}(x)=\sqrt{\frac{\beta }{2^{n}n!\sqrt{\pi }}}H_{n}(\beta x)\exp (\frac{%
-\beta ^{2}x^{2}}{2}),$\ where, $H_{n}(\beta x)$ are Hermite polynomials and 
$\beta =(\frac{2\sqrt{V_{0}}}{k^{\hspace{-2.1mm}-}})^{\frac{1}{2}}=\sqrt{2}%
q^{\frac{1}{4}}.$

The time evolution of the particle, initially in state $\psi (x,0),$ is
obtained by time evolution operator $\hat{U},$ such that, $\psi (x,t)=\hat{U}%
\psi (x,0)=\sum\limits_{n=0}^{\infty }c_{n}\phi _{n}(x)\exp (-i\frac{E_{n}}{%
k^{\hspace{-2.1mm}-}}t),$\ where, $E_{n}$ and $\phi _{n}(x)$ are energy
eigen values and eigen states corresponding to quantum number, $n.$ The
probability amplitudes $c_{n}$ are defined as, $\langle \phi _{n}(x)|\psi
(x,0)\rangle .$ The quantum particle wave packet in optical potential,
narrowly peaked around a mean quantum number $\bar{n},$ displays quantum
recurrences at different time scales, defined as$\ T_{(j)}=\frac{2\pi }{%
(j!k^{\hspace{-2.1mm}-})^{-1}E_{n}^{j}|_{n=\bar{n}}},$ where, $E_{n}^{j}$
denotes the $j^{th}$ derivative of $E_{n}$ with respect to $n$. The time
scale, $T_{(1)},$ is termed as classical period as it provides a time at
which the particle completes its evolution following the classical
trajectory and reshapes itself. Whereas, at$\ T_{(2)}$ the particle reshapes
itself as a consequence of quantum interference in a nonlinear energy
spectrum, which is purely a quantum phenomenon and thus named as quantum
revival time. In the parametric regime of a changing nonlinearity with
respect to quantum number, $n,$ we find, the super revival time $T_{(3)}$
for the quantum particle \cite{LeichtleAverbukhSchleich}.

We study the time evolution of the material wave packet using square of the
autocorrelation function \cite{Nauenberg}\textit{,}%
\begin{equation*}
|A(t)|^{2}=\sum\limits_{n=0}^{\infty }|c_{n}|^{4}+2\sum\limits_{n\neq
m}^{\infty }|c_{n}|^{2}|c_{m}|^{2}\cos [(E_{n}-E_{m})\frac{t}{k^{\hspace{%
-2.1mm}-}}].
\end{equation*}%
In the present parametric regime the square of the autocorrelation function
is written as%
\begin{equation}
|A(t)|^{2}=\sum\limits_{n=0}^{\infty }|c_{n}|^{4}+2\sum\limits_{n\neq
m}^{\infty }|c_{n}|^{2}|c_{m}|^{2}\cos ((n-m)2\sqrt{V_{0}}t),
\label{autofun}
\end{equation}%
where, $\sum\limits_{n=0}^{\infty }|c_{n}|^{4}$\ is independent of time, and
defines the interference free, averaged value of $|A(t)|^{2}$. The most
dominant contribution to the $|A(t)|^{2}$\ comes from $n-m=1$\ in the second
part at the right side of equation \ref{autofun}. Other terms with, $m-n\geq
1,$\ have negligible role because their oscillation frequency is an integral
multiple of fundamental frequency $2\sqrt{V_{0}},$\ and are averaged out to
zero. For the reason the square of the autocorrelation function in the
present case oscillates following a cosine law with a frequency $2\sqrt{V_{0}%
}$\ which leads to a classical time period $T_{cl}^{(0)}=\frac{\pi }{\sqrt{%
V_{0}}}$ as shown in figure 2. Here, zero in the superscript of $%
T_{cl}^{(0)} $ defines the system's classical period in the absence of
perturbation. At the integral multiples of classical period, $|A(t)|^{2}$\
is unity, whereas, at time, which is odd integral multiple of the half of
the classical period%
\begin{equation*}
|A(t)|^{2}=\sum\limits_{n=0}^{\infty }|c_{n}|^{4}+2\sum\limits_{n\neq
m}^{\infty }|c_{n}|^{2}|c_{m}|^{2}\cos [(n-m)\pi ].
\end{equation*}%
Here $\cos [(n-m)\pi ]\ $has alternatively values $+1$\ and $-1,$\ when $n-m$%
\ is an even or odd respectively. Thus after the cancellation of positive
terms with the negative ones, second summation reduces to minimum value and $%
|A(t)|^{2}$\ attains its minima. Here, in the eigen states expansion%
\begin{equation}
\psi (x,t)=\exp [-i(2\sqrt{V_{0}}-\frac{V_{0}}{k^{\hspace{-2.1mm}-}}%
)t]\sum\limits_{n=0}^{\infty }c_{n}\phi _{n}(x)\exp (-i2n\sqrt{V_{0}}t),
\end{equation}%
it is notable that the eigen states $\phi _{n}(x)$ have parity $(-1)^{n}.$
In case of even parity, only the even terms $c_{2n}$ are nonvanishing and $n$%
-dependent exponent factor oscillates two times faster than the general
case. At half of classical period the wave packet reappears towards other
turning point of the well. In case the wave packet is initially placed at
the center of the cosine well, it reappears at half classical period at the
same position but in opposite direction. In this case we see classical
revivals of the initial atomic wavepacket and quantum revivals take place at
infinitely long time. Hence, we find revival of the atomic wavepacket after
each classical period only. Experimentally we may realize the situation by
placing very few recoil energy atoms deep in the cosine potential well. This
reveals the information about the level spacing around the bottom of the
cosine potential. Interestingly we find an equal spacing between the energy
levels from figure 1, for large $q$ and small value of $n.$ The
spatiotemporal behavior of \ the wave packet in the quadratic potential, as
shown in figure 2, confirms the above results.

\subsection{\textbf{Quartic Oscillator Limit.}}

Beyond harmonic oscillator limit, we find oscillator with nonlinearity and
energy level spacing different from a constant value. The correction to the
energy of harmonic oscillator comes from the first order perturbation (see
appendix for energy corrections), that is, $E_{n}^{(4)}=-\frac{k^{\hspace{%
-2.1mm}-2}}{8}(2n^{2}+2n+1),$ which again can be identified in equation \ref%
{eq3}\ by ignoring cubic and higher order powers in $n.$\ The atoms with
little higher energy, which is equivalent to several recoil energies observe
another time scale in which it reconstructs itself beyond classical period,
i.e. quantum revival time. The behavior of auto-correlation function for the
wavepacket exactly placed in this region, where only first order correction
is sufficient, is shown in figure 3. We see that the wave packet displays
revivals at quantum revival time. Thus, little above from the bottom of the
of an optical lattice, the wave packet sees variations in energy level
spacing in a non-linear fashion. The classical time is modified as $%
T_{cl}^{(1)}=\alpha ^{(1)}T_{cl}^{(0)},$ where $\alpha ^{(1)}=1+\frac{\bar{s}%
}{8\sqrt{q}}$, and classical periodicity for a particle in present situation
is related to potential height and mean quantum number, here, $\bar{s}=2\bar{%
n}+1.$ As $\bar{n}$ increases, classical revival time also increases, and
the ratio, $\frac{\bar{s}}{8\sqrt{q}},$ is always less than unity in the
region where first correction is sufficient. The quantum revival time, $%
T_{rev}^{(1)}=\frac{8\pi }{k^{\hspace{-2.1mm}-}},$ is independent of mean
quantum number $\bar{n},$ whereas super revival time in this case is
infinite.

The eigen states of quartic oscillator due to first order perturbation are
given as $\phi _{n}^{q}(x)=\phi _{n}(x)+\phi _{n}^{(1a)}(x),$where, $\phi
_{n}^{(1a)}(x)$ is first order correction to the harmonic oscillator wave
function and is defined as%
\begin{equation}
\phi _{n}^{(1a)}=D_{1}(\eta _{1}\phi _{n-4}+\eta _{2}\phi _{n-2}-\eta
_{3}\phi _{n+2}-\eta _{4}\phi _{n+4}).
\end{equation}%
Here, $D_{1},$ $\eta _{1},$ $\eta _{2},$ $\eta _{3}$ and $\eta _{4}$ are
constants and defined in appendix.

In figure 4 eigen states of quadratic, quartic, sixtic and octic oscillators
are mapped on numerically obtained eigen states of cosine potential. From
this figure we see that for $V_{0}=10$\ and $k^{\hspace{-2.1mm}-}=0.5,$
first order correction to the eigen states of unperturbed system matches
with harmonic oscillator eigen states up to $n=3,$\ and mapping of quartic
oscillator with exact solution is much improved compare to harmonic
oscillator. Similarly mapping of sixtic oscillator is better than quartic
oscillator and is quite good for octic oscillator for all bound band. In
this case eight bands exist in side the potential. From figure 1 we note
that by increasing $q$ number of bound bands can be increased.

The square of autocorrelation function in this case takes the form%
\begin{eqnarray}
|A(t)|^{2} &=&\sum\limits_{n=0}^{\infty }|c_{n}|^{4}+2\sum\limits_{n\neq
m}^{\infty }|c_{n}|^{2}|c_{m}|^{2}\cos [(n-m)2\sqrt{V_{0}}t 
+(m-n)(n+m+1)\frac{k^{\hspace{-2.1mm}-}}{4}t].  \label{AutoRev}
\end{eqnarray}%
Here, nonlinear dependence of energy eigen values on quantum number, $n,$
makes the argument of cosine function non-linear, as appears in the last
term of the above expression. The nonlinear term $(m-n)(n+m+1)$ removes a
degeneracy present for the harmonic case between the cosine waves
corresponding to nearest neighbouring off diagonal terms and beyond. Hence,
there overall evolution display a gradual decoherence leading to collapse
which latter transforms in revival as the decoherence in waves disappears.\
The values of $|A(t)|^{2}$ in this regime at $T_{rev}=\frac{8\pi }{k^{%
\hspace{-2.1mm}-}}$ is simplifies as%
\begin{equation}
|A(t)|^{2}=\sum\limits_{n=0}^{\infty }|c_{n}|^{4}+2\sum\limits_{n\neq
m}^{\infty }|c_{n}|^{2}|c_{m}|^{2}\cos [16\pi \sqrt{q}(n-m)].
\label{RevTime}
\end{equation}%
The equation \ref{RevTime} displays that $|A(t)|^{2}=1,$ when $\sqrt{q}$ is
an integral multiple of $\frac{1}{8}$\ and also $|A(t)|^{2}$ is unity at
half of the revival time if $\sqrt{q}$\ is an integral multiple of $\frac{1}{%
4}$. In case the $\sqrt{q}$ is not an integral multiple of $\frac{1}{8}$,
the wave packet revival occurs little earlier than the revival time $T_{rev}=%
\frac{8\pi }{k^{\hspace{-2.1mm}-}}$ and also $|A(t)|^{2}$ approaches to
unity little earlier than $\frac{T_{rev}}{2}$. Here reproduction of wave
packet at $\frac{T_{rev}}{2}$ is out of phase by $\pi ,$ i.e. at that time
all the waves are moving exactly in opposite directions as initially they
were. But at $T_{rev}$ they are all moving in same direction and each wave
is in phase not only with there initial states but also with each other.

Figure 5 shows spatiotemporal behavior of the wave packet in cosine
potential. We see that wave packet spreads and oscillates in the cosine
well, and after some time the original wavepacket is divided into
sub-wave-packets figure 5(b). At a quantum revival time these
sub-wave-packets constructively interfere and wave packet gains the original
shape at the same initial position figure 5(c). At quantum revival time,
same classical pattern is seen as at the start of the time evolution shown
in figure 5(a).

\subsection{\textbf{Sixtic Oscillator Limit: Existence of Super Revivals and
Beyond}}

Higher order nonlinearity in the energy spectrum of the quantum pendulum
show up beyond quartic limit. In the presence of the second correction, the
energy is modified by the term $E_{n}^{(6)}=-\frac{k^{\hspace{-2.1mm}-3}}{%
\sqrt{V_{0}}}\frac{(2n^{3}+3n^{2}+3n+1)}{32}.$ Second correction to the
energy modifies the time scales. The classical time period is now $%
T_{cl}^{(2)}=\alpha ^{(2)}T_{cl}^{(0)},$ whereas the quantum revival time is
modified as $T_{rev}^{(2)}=|\beta ^{(1)}|T_{rev}^{(1)}.$ Here, $\alpha
^{(2)}=\alpha ^{(1)}+\frac{3(\bar{s}^{2}+1)}{2^{8}q},$ $\beta ^{(1)}=\frac{3%
\bar{s}}{16q}-1$ are constants. In addition system shows another time scale,
i.e. super revival time $T_{spr}^{(2)}=\frac{64\pi \sqrt{V_{0}}}{k^{\hspace{%
-2.1mm}-2}},$ at which reconstruction of original wave packet takes place.
The other quantum revival times occur at infinity.

The energy eigen states in this regime are given as $\phi _{n}^{(s)}=\phi
_{n}+\phi _{n}^{(1,a)}+\phi _{n}^{\left( 1,b\right) }+\phi _{n}^{\left(
2,a\right) },$ where, $\phi _{n}^{\left( 2,a\right) }$ is perturbation in
eigen states due to the $H^{(6)}$ term and $\phi _{n}^{\left( 1,b\right) }$
is second order perturbation caused by $H^{(4)}$ term \cite{Doncheski}. The
expressions $\phi _{n}^{\left( 1,b\right) }$ and $\phi _{n}^{\left(
2,a\right) }$ are given as%
\begin{eqnarray*}
\phi _{n}^{\left( 1,b\right) } &=&D_{2}[\delta _{1}\phi _{n-8}+\delta
_{2}\phi _{n-6}+\delta _{3}\phi _{n-4}+\delta _{4}\phi _{n-2} +\delta _{5}\phi
_{n+2}+\delta _{6}\phi _{n+4}+\delta _{7}\phi
_{n+6}+\delta _{8}\phi _{n+8}],
\end{eqnarray*}%
and%
\begin{eqnarray*}
\phi _{n}^{\left( 2,a\right) } &=&D_{6}[\chi _{_{1}}\phi _{n-6}+\chi
_{_{2}}\phi _{n-4}+\chi _{_{3}}\phi _{n-2}+\chi _{_{4}}\phi _{n+2} +\chi
_{_{5}}\phi _{n+4}+\chi _{_{6}}\phi _{n+6}].
\end{eqnarray*}%
The constants $D_{2},D_{6},\delta _{j}$'$s$ and $\chi _{j}$'$s$ are
calculated in appendix, where, $j$ takes integral values.

Again in this region classical time increases by increasing $\bar{n}$ and
increases faster than as it was in quartic limit. However, in the present
regime, the quantum revival time is not constant and decreases as $\bar{n}$
increases. The $\bar{n}$ dependence of quantum revival time is shown in
figure 6. The super revival time is independent of mean quantum number. It
is directly proportional to square root of potential height and inversely
proportional to the square of scaled Planck's constant. The temporal
behavior of an atom in optical potential placed in this regime shows three
time scales: classical periods enveloped in quantum revivals and quantum
revivals enveloped in super revivals as shown in figure 7. After each super
revival time the atomic wave packet evolution repeats itself.

Similarly third correction in energy, modify the energy by the factor $%
E_{n}^{(8)}=-\frac{k^{\hspace{-2.1mm}-4}}{V_{0}}\frac{%
(5n^{4}+10n^{3}+16n^{2}+11n+3)}{2^{8}}$. The time scales in this case are $%
T_{cl}^{(3)}=\alpha ^{(3)}T_{cl}^{(0)}$ and$\ T_{rev}^{(3)}=|\beta
^{(2)}|T_{rev}^{(1)}$ Where $\alpha ^{(3)}=\alpha ^{(2)}+\frac{(5\bar{s}%
^{3}+17\bar{s})}{2^{11}q^{\frac{3}{2}}}\ $and $\beta ^{(2)}=\beta ^{(1)}+%
\frac{15\bar{s}^{2}+17}{2^{8}q}$ and super revival time is $%
T_{spr}^{(3)}=|\gamma |T_{spr}^{(2)}$ where $\gamma =\frac{5\bar{s}}{8\sqrt{q%
}}-1.$ Furthermore, the super quartic revival time, $T_{4}$, is independent
of $\bar{n}$. The higher order corrections in energy show that other time
scales do exist in the system, but their times of recurrence are too large
to consider them finite.

In the presence of third correction to energy, classical time increases as $%
\bar{n}$ increases but increases little faster than in the cases of quartic
and sixtic corrections, whereas, the quantum revival time decreases faster
as $\bar{n}$ increases compared to the case of sixtic correction. The super
revival time is not constant but decreases as $\bar{n}$ increases.

Energy corrections increase anharmonicity in the system. We discussed that
the first order correction to harmonic potential energy led to the quantum
revivals, the second order energy correction led to super revivals and
fourth correction led to quartic revival time. Comparison of revival times
for different energy corrections is shown in figure 6. For first order
energy correction quantum\ revival time is constant, but for higher order
corrections, revival time may decrease with increasing mean quantum number
of the wavepacket.

In figure 4\ we\ show the projection of numerically calculated eigen states
of cosine potential on the eigen states of quadratic, quartic (first
correction to quadratic potential), sixtic (second correction to quadratic
potential) and octic (third correction to quadratic potential) potentials.
We note that the eigen states of all above mentioned potentials match to the
eigen states of quadratic potential near the bottom of lattice potential.
For little large quantum numbers, we see that projection of quadratic
potential falls sharply and improves with higher order potentials. This
correction is quite good for octic potential when $q=40$ justifying $q>>1$
condition. Eigen states and eigen energies of quartic, sixtic and octic
potentials are analytically calculated using the method given in appendix.

\section{Discussion}

In this paper we have extended\ the understanding of eigen energy levels and
eigen states in deep optical lattice, both analytically and numerically. We
note that solutions obtained through perturbation theory and Mathieu
solution are showing similar results and are in very good agreement with
exact numerical solutions. The energy levels are equally spaced near the
bottom and by increasing the lattice potential depth, the number of equally
spaced energy levels can be increased. A wave packet placed in this region
revives after each classical period. From figure 4, it is also noted that
all potentials discussed have equally spaced eigen levels at the bottom of
the potential well as their mapping with harmonic oscillator is unity.
Beyond linear regime, energy dependence is quadratic and any wavepacket
evolved in this region shows complete quantum revivals enveloped by
classical revivals. Interestingly in this regime quantum revival time ($%
T_{rev}^{(1)}=\frac{8\pi }{k^{\hspace{-2.1mm}-}}$) is independent of
potential height, however has inverse proportionality with effective
Planck's constant, $k^{\hspace{-2.1mm}-}$. We show that for deep optical
lattice, there is a region in which revivals are independent of lattice
depth and this region expands with the increase in lattice depth, however
beyond this region quantum revival time is no longer constant but decreases
as $\bar{n}$ increases. The higher order time scale, super revival time $%
T_{spr}^{(2)}=\frac{64\pi \sqrt{V_{0}}}{k^{\hspace{-2.1mm}-2}}$ exists in
this region and is independent of $\bar{n}$. Again this region expands as
potential height is increased but super revival time in this region is
directly proportional to square root of potential height.

Above this region other time scales also exist where super revival time is $%
\bar{n}$ dependent and decreases as $\bar{n}$ increases but these time
scales are too long to consider them finite.

\section{Acknowledgement}

M. A. and K. N. thank Higher Education Commission Pakistan for financial
support through grant No.17-1-1(Q.A.U)HEC/Sch/2004/5681. FS is supported by
Higher Education Commission Pakistan through research grant 20-23 R \&
D/03143, The Abdus Slam International Center for Theoretical Physics, ICTP,
Trieste, Italy and Pakistan Science Foundation. He thanks S.\ Stanislav and
G. Ghirardi for fruitful discussions. We thank S. Iqbal, Rameez-ul-Islam, I.
Rehman and T.\ Abbas for useful suggestions.

\bigskip

\section{Appendix: Solution of an Arbitrary Potential}

An arbitrary potential $U(r),$ around its minima can be solved by taking its
Taylor's expansion \cite{Liboff}, that is%
\begin{eqnarray}
U(r) &=&U(r_{m})+G^{(1)}(r-r_{m})+G^{(2)}(r-r_{m})^{2}
+G^{(3)}(r-r_{m})^{3}+....,  \label{expansion}
\end{eqnarray}%
where, $G^{(j)}=(j!)^{-1}\partial ^{j}U(r=r_{m})/\partial r^{j}$, and $j$ is
an integer. The value of $G^{(j)}$ for odd $j$\ is zero as potential is $%
\cos (x)$ and it is calculated at the potential minima $r=r_{m}$. Thus in
the presence of weak nonlinearity the term $G^{(6)}<<G^{(4)}<<G^{(2)}$.

In our analysis we consider the expansion up to second order term\ as
unperturbed Hamiltonian, $H_{0}$. The eigen functions and eigen energies of
this Hamiltonian are those of harmonic oscillator. The effect of the higher
order terms in Taylor's expansion is discussed as perturbation to eigen
energies and eigen functions of the\ harmonic oscillator. We express the
effective Hamiltonian governing the dynamics of atom around the potential
minima as%
\begin{equation}
H_{0}\cong \frac{{\hat{p}}^{2}}{2}+U(r_{m})+G^{(2)}(r-r_{m})^{2}.
\label{ham1}
\end{equation}%
The eigen functions and eigen energies of the harmonic potential are $\phi
_{n}(x)=\sqrt{\frac{\beta }{2^{n}n!\sqrt{\pi }}}H_{n}(\beta x)\exp (\frac{%
-\beta ^{2}x^{2}}{2}),$ and $E_{n}^{(0)}=k^{\hspace{-2.1mm}-}\sqrt{V_{0}}%
(2n+1)+U(r_{m}),$ where, $H_{n}(\beta x)$ are Hermite polynomials.

The first order correction and second order correction to energy is quite
well known. The first order correction to the energy of quadratic potential
is $E_{n}^{(p,a)}=\langle \phi _{n}|H^{(p)}|\phi _{n}\rangle ,$ and the
second order correction is obtained from the relation $E_{n}^{(p,b)}=\sum_{m%
\neq n}\frac{|\langle \phi _{n}|H^{(p)}|\phi _{m}\rangle |^{2}}{%
E_{n}^{(0)}-E_{m}^{(0)}}.$\ The third order correction is determined by%
\begin{eqnarray*}
E_{n}^{(p,c)} &=&\sum_{j\neq n}\sum_{l\neq n}\frac{\langle n|\hat{H}%
^{(p)}|j\rangle \langle j|\hat{H}^{(p)}|l\rangle \langle l|\hat{H}%
^{(p)}|n\rangle }{(E_{n}^{(0)}-E_{j}^{(0)})(E_{n}^{(0)}-E_{l}^{(0)})}-\langle
n|\hat{H}^{(p)}|n\rangle \sum_{j\neq n}\frac{\langle n|\hat{H}%
^{(p)}|j\rangle \langle j|\hat{H}^{(p)}|n\rangle }{%
((E_{n}^{(0)})^{2}-(E_{j}^{(0)})^{2})},
\end{eqnarray*}%
where,$\ p=3,4,5,6,........\cdot $\bigskip

The presence of the perturbation term $H^{(4)},$ leads to the Hamiltonian 
\begin{equation}
H=H_{0}+H^{(4)}.  \label{tham}
\end{equation}%
Here, the first order correction to eigen functions, due to the $H^{(4)}$
term, is obtained as 
\begin{equation}
|\phi _{n}^{(1a)}\rangle =D_{1}(\eta _{1}\phi _{n-4}+\eta _{2}\phi
_{n-2}-\eta _{3}\phi _{n+2}-\eta _{4}\phi _{n+4}),  \label{ef2}
\end{equation}%
where, $D_{1}=G^{(4)}(\frac{1}{4\sqrt{q}})^{2}\frac{1}{k^{\hspace{-2.1mm}%
-}\omega _{h}},$ 
\begin{align*}
\eta _{1}& =\sqrt{n(n-1)(n-2)(n-3)}/4, \\
\eta _{2}& =(2n-1)\sqrt{n(n-1)},\text{ } \\
\eta _{3}& =(2n+3)\sqrt{(n+1)(n+2)}, \\
\text{and }\eta _{4}& =\sqrt{(n+1)(n+2)(n+3)(n+4)}/4.
\end{align*}

Now the eigen functions in the presence of first order perturbation, due to
the correction $H^{(4)},$\ is given as%
\begin{equation*}
\phi _{n}^{q}(x)=\phi _{n}(x)+\phi _{n}^{(1a)}(x),
\end{equation*}%
and second order correction due to $H^{\left( 4\right) }$ term is 
\begin{align}
\phi _{n}^{\left( 1,b\right) }& =D_{2}[\delta _{1}\phi _{n-8}+\delta
_{2}\phi _{n-6}+\delta _{3}\phi _{n-4}+\delta _{4}\phi _{n-2}   +\delta
_{5}\phi _{n+2}+\delta _{6}\phi _{n+4}+\delta _{7}\phi
_{n+6}+\delta _{8}\phi _{n+8}.
\end{align}%
Here, $D_{2}=(G^{\left( 4\right) })^{2}(\frac{1}{4\sqrt{q}})^{4}(\frac{1}{k^{%
\hspace{-2.1mm}-}\omega _{h}})^{2},$ 
\begin{eqnarray*}
\delta _{1} &=&\frac{\sqrt{n\left( n-1\right) \left( n-2\right) \left(
n-3\right) \left( n-4\right) \left( n-5\right) \left( n-6\right) \left(
n-7\right) }}{32}, \\
\delta _{2} &=&\frac{\left( 6n-11\right) }{12}\sqrt{n\left( n-1\right)
\left( n-2\right) \left( n-3\right) \left( n-4\right) \left( n-5\right) }, \\
\delta _{3} &=&\left( 2n^{2}-9n+7\right) \sqrt{n\left( n-1\right) \left(
n-2\right) \left( n-3\right) }, \\
\delta _{4} &=&\frac{\left( 56n^{3}-228n^{2}+214n-146\right) }{8}\sqrt{%
n\left( n-1\right) }, \\
\delta _{5} &=&\frac{\left( 56n^{3}+396n^{2}+838n+645\right) }{8}\sqrt{%
\left( n+1\right) \left( n+2\right) }, \\
\delta _{6} &=&\frac{(31n^{2}+197n+258)}{16}\sqrt{\left( n+1\right) \left(
n+2\right) \left( n+3\right) \left( n+4\right) }, \\
\delta _{7} &=&\tfrac{\left( 11n+27\right) }{24}\sqrt{\left( n+1\right)
\left( n+2\right) \left( n+3\right) \left( n+4\right) \left( n+5\right)
\left( n+6\right) }, \\
\text{and }\delta _{8} &=&\tfrac{\sqrt{\left( n+1\right) \left( n+2\right)
\left( n+3\right) \left( n+4\right) \left( n+5\right) \left( n+6\right)
\left( n+7\right) \left( n+8\right) }}{32}.
\end{eqnarray*}

Hence, following the same procedure, the first order correction due to $%
H^{\left( 6\right) }$ term changes the Hamiltonian of the system as%
\begin{equation}
H=H_{0}+H^{\left( 4\right) }+H^{\left( 6\right) }.
\end{equation}%
Hence the corrected eigen function in presence of correction due to $H^{(4)}$
and $H^{\left( 6\right) }$ terms appear as%
\begin{equation*}
\phi _{n}^{(s)}=\phi _{n}+\phi _{n}^{(1,a)}+\phi _{n}^{\left( 1,b\right)
}+\phi _{n}^{\left( 2,a\right) },
\end{equation*}%
where,%
\begin{align*}
\phi _{n}^{\left( 2,a\right) }& =D_{6}[\chi _{_{1}}\phi _{n-6}+\chi
_{_{2}}\phi _{n-4}+\chi _{_{3}}\phi _{n-2} +\chi _{_{4}}\phi _{n+2}+\chi
_{_{5}}\phi _{n+4}+\chi _{_{6}}\phi _{n+6}.
\end{align*}%
Here, $D_{6}=G^{(6)}(\frac{1}{4\sqrt{q}})^{3}\frac{1}{k^{\hspace{-2.1mm}%
-}\omega _{h}},$%
\begin{eqnarray*}
\chi _{_{1}} &=&6\sqrt{n(n-1)(n-2)(n-3)(n-4)(n-5)}, \\
\chi _{_{2}} &=&\frac{3}{4}(2n-3)\sqrt{n(n-1)(n-2)(n-3)}, \\
\chi _{_{3}} &=&\frac{15}{2}(n^{2}-n+1)\sqrt{n(n-1)}, \\
\chi _{_{4}} &=&\frac{15}{2}(n^{2}+3n+3)\sqrt{(n+1)(n+2)}, \\
\chi _{_{5}} &=&\frac{3}{4}(2n+5)\sqrt{\left( n+1\right) \left( n+2\right)
\left( n+3\right) \left( n+4\right) }, \\
\chi _{_{6}} &=&2\sqrt{\left( n+1\right) \left( n+2\right) \left( n+3\right)
\left( n+4\right) \left( n+5\right) \left( n+6\right) }.
\end{eqnarray*}

Close to the minima of potential, we can find the eigen energies up to a
considerable accuracy by using perturbation theory. The leading correction
comes from $H^{(4)}$ using first order and second order perturbation theory
respectively. The result is as under:%
\begin{equation*}
E_{n}^{(4)}=(\alpha _{2}n^{2}+\alpha _{1}n+\alpha _{0})k^{\hspace{-2.1mm}%
-}\omega _{h},
\end{equation*}%
where, $\alpha _{0}=3D_{a},$ $\alpha _{1}=6D_{a},$ $\alpha _{2}=6D_{a},$ $%
C_{b=}(G^{(3)})^{2}(\frac{1}{4\sqrt{q}})^{3}\frac{1}{k^{\hspace{-2.1mm}%
-}\omega _{h}}$ and $D_{a}=G^{(4)}(\frac{1}{4\sqrt{q}})^{2}.$ At next order,
the first order perturbation of $H^{(6)}$and second order perturbation of $%
H^{(4)}$ contribute. The result can be written as: 
\begin{equation*}
E_{n}^{(6)}=(\beta _{3}n^{3}+\beta _{2}n^{2}+\beta _{1}n+\beta _{0})k^{%
\hspace{-2.1mm}-}\omega _{h},
\end{equation*}%
where, $\beta _{0}=3I_{a}-21J_{b},\ \beta _{1}=8I_{a}-59J_{b},$ $\beta
_{2}=6I_{a}-51J_{b},$ $\beta _{3}=4I_{a}-34J_{b},$ and $I_{a}=5G^{(6)}(\frac{%
1}{4\sqrt{q}})^{3}\frac{1}{k^{\hspace{-2.1mm}-}\omega _{h}},$ $%
J_{b}=2(G^{(4)})^{2}(\frac{1}{4\sqrt{q}})^{2}(\frac{1}{k^{\hspace{-2.1mm}%
-}\omega _{h}})^{2}.$

At the next higher order, we need to evaluate three contributions; $H^{(8)}$
in first order, $H^{(6)}$ and $H^{(4)}$ in second order and $H^{(4)}$in
third order. The energy expression is%
\begin{equation*}
E_{n}^{(8)}=(\gamma _{4}n^{4}+\gamma _{3}n^{3}+\gamma _{2}n^{2}+\gamma
_{1}n+\gamma _{0})k^{\hspace{-2.1mm}-}\omega _{h},
\end{equation*}%
where, $\gamma _{0}=3X-12Y-111Z,$ $\gamma _{1}=8X-35Y-347Z,$ $\gamma
_{2}=10X-46Y-472Z,$ $\gamma _{3}=4X-22Y-250Z,$ $\gamma _{4}=2X-11Y-125Z,$
and $X=35G^{(8)}(\frac{1}{4\sqrt{q}})^{4}\frac{1}{k^{\hspace{-2.1mm}-}\omega
_{h}},Y=30G^{(6)}G^{(4)}(\frac{1}{4\sqrt{q}})^{5}(\frac{1}{k^{\hspace{-2.1mm}%
-}\omega _{h}})^{2},Z=48(G^{(4)})^{3}(\frac{1}{4\sqrt{q}})^{6}(\frac{1}{k^{%
\hspace{-2.1mm}-}\omega _{h}})^{2}.$

Now the energy of the system is%
\begin{eqnarray}
{E}_{n} &=&E_{n}^{(0)}+E_{n}^{(4)}+E_{n}^{(6)}+E_{n}^{(8)},\text{ or } 
\notag \\
E_{n} &=&(\kappa _{4}n^{4}+\kappa _{3}n^{3}+\kappa _{2}n^{2}+\kappa
_{1}n+\kappa _{0})k^{\hspace{-2.1mm}-}\omega _{h}+U(r_{m}),  \label{ee}
\end{eqnarray}%
where, $\kappa _{0}=\alpha _{0}+\beta _{0}+\gamma _{0}+\frac{1}{2},$ $\kappa
_{1}=\alpha _{1}+\beta _{1}+\gamma _{1}+1,$\ $\kappa _{2}=\alpha _{2}+\beta
_{2}+\gamma _{2},$\ $\kappa _{3}=\beta _{3}+\gamma _{3},$ and $\kappa
_{4}=\gamma _{4}.$

\begin{figure}[h]
\begin{center}
\includegraphics[width=12cm,height=10cm]{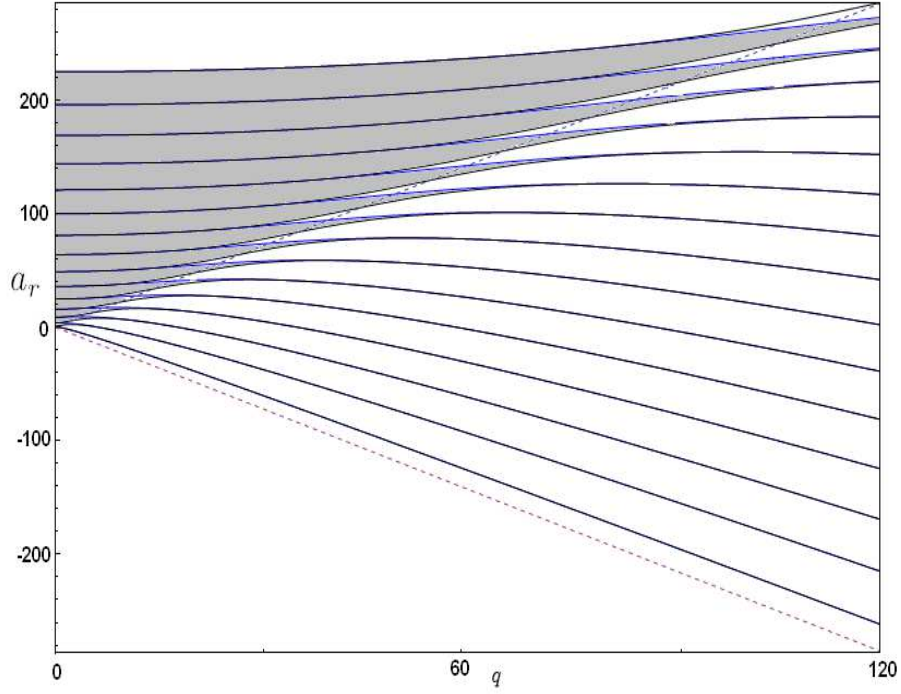}
\end{center}
\caption{Characteristic values for the Mathieu equation, $a_{2m}$ (for even
solutions, blue (grey) curves) and $b_{2m}$ (for odd solutions, red (dark)
curves, mostly overridden by blue curves) versus $q$ for the quantum
pendulum: The dotted lines correspond to characteristic parameter $a_{n}=\pm
2q$ i.e. $E=\pm V_{0}.$ Note that for a given value of $q$ where $q\gg a_{n},
$ the gap between the lowest energy state (lowest solid curve) is roughly
one-half of the spacing between solid and dashed curves, corresponding to
the zero-point energy in the oscillator limit.}
\label{Fig-1}
\end{figure}

\begin{figure}[h]
\begin{center}
\includegraphics[width=12cm,height=4cm]{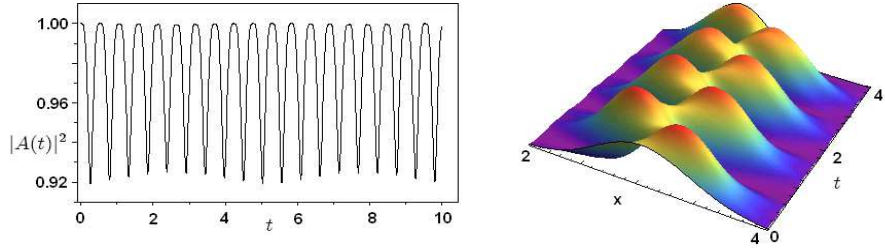}
\end{center}
\caption{Time evolution of particle wave packet placed at the bottom of the
cosine \ potential. The dimensions of the wave packet are $k^{\hspace{-2.1mm}%
-}=0.5,$ $\Delta p=0.5$ with $V_{0}=10.$ We show autocorrelation function vs
time (right side) and spatiotemporal behavior of the material wavepacket
(left side). The wave packet see equally spaced energy levels and rebuilds
after every classical period. Analytically calculated value of classical
period and numerical results show an excellent agreement.}
\label{Fig-2}
\end{figure}

\begin{figure}[h]
\begin{center}
\includegraphics[width=12cm,height=10cm]{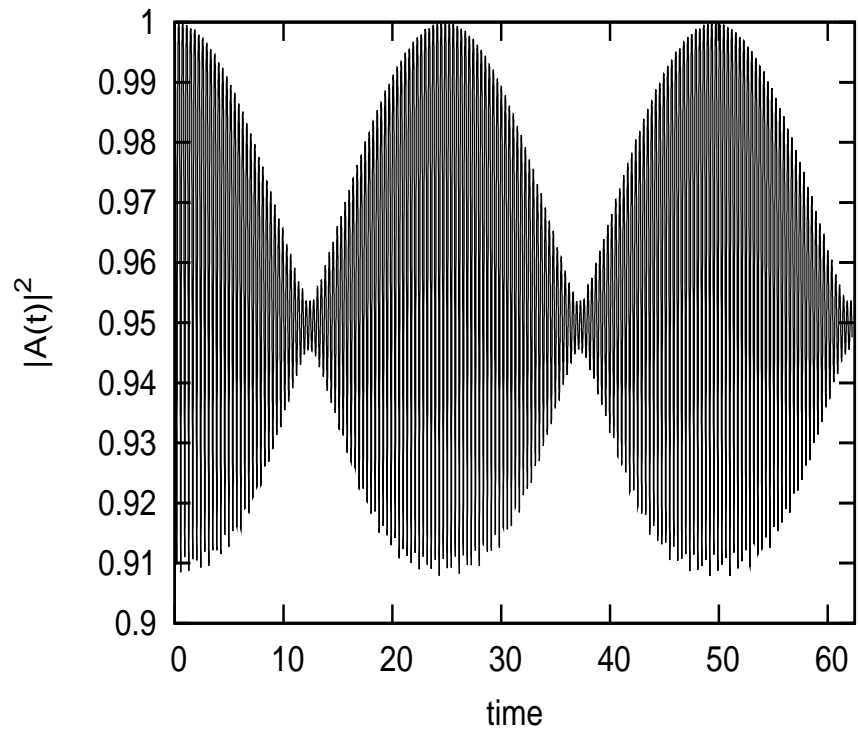}
\end{center}
\caption{Autocorrelation function for a particle undergoing quantum revival
evolution in time. The parameters are same as in figure 2. The wave packet
was placed close to the bottom of the potential well in the regime where
first order correction is sufficient, it observes quantum revivals after
many classical periods.}
\label{Fig-3}
\end{figure}

\begin{figure}[h]
\begin{center}
\includegraphics[width=12cm,height=10cm]{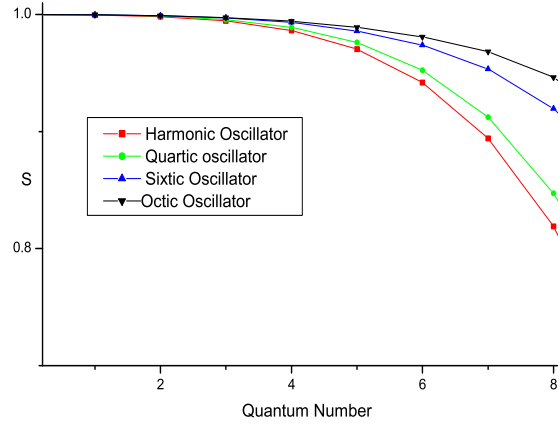}
\end{center}
\caption{comparison of cosine potential with the simplified potentials is
made by calculating the projection, S, of the eigen states of the cosine
potential on the eigen states of the simplified potentials . For first few
quantum numbers, the cosine potential very much resembles to the harmonic
potential, however, for higher quantum numbers the higher order corrections
to the harmonic oscillator are needed to make the resemblance.
}
\label{Fig-4}
\end{figure}

\begin{figure}[h]
\begin{center}
\includegraphics[width=12cm,height=4cm]{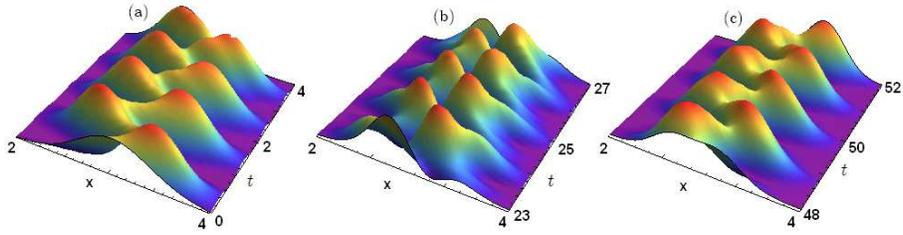}
\end{center}
\caption{The time evolution of a localized wave packet in a cosine potential
is displayed for the same parameters as in figure 2. For short times the
initial wavepacket shows classical revivals (a), but starts to display
sub-wave-packets in its long time dynamics (b), which constructively
interfere at quantum revival time $T_{rev}$ (c). Our analytical and
numerical results are in very good agreement.}
\label{Fig-5}
\end{figure}

\begin{figure}[h]
\begin{center}
\includegraphics[width=12cm,height=10cm]{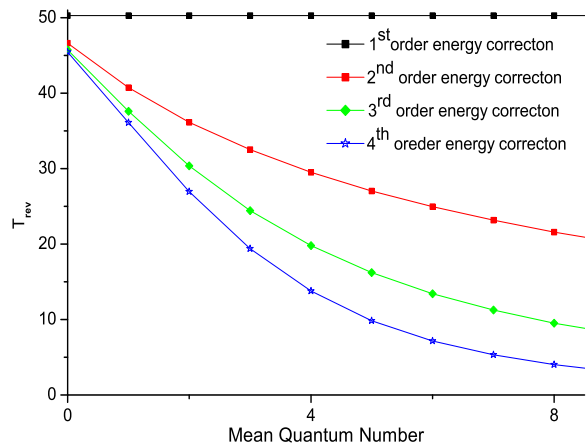}
\end{center}
\caption{The quantum revival time vs mean quantum number is shown for
simplified potentials. In the presence of only first order correction, i.e.
for quartic potential, the quantum revival time is a constant. For higher
order corrections it decreases with increasing mean quantum number.}
\label{Fig-6}
\end{figure}

\begin{figure}[h]
\begin{center}
\includegraphics[width=11cm,height=8cm]{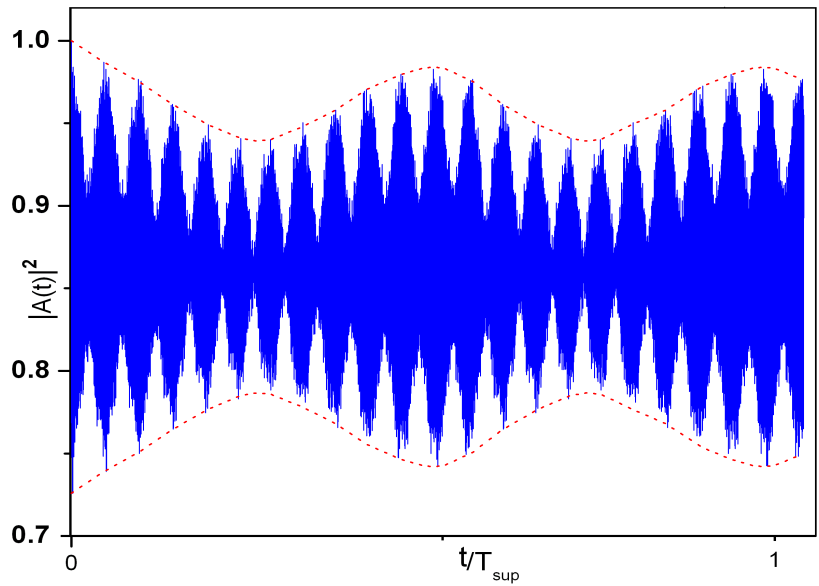}
\end{center}
\caption{The wave packet dynamics in sixtic potential displays three time
scales, the classical periods (making the dense region), the quantum revival
times (making the peaks in the dense region) and the super revival times
(making the peaks in the envelop of the quantum revivals). The parametric
values are the same as in figure 2.}
\label{Fig-7}
\end{figure}

\end{document}